\documentclass[11pt]{article}
\usepackage{epsf}
\usepackage{amsmath}
\usepackage{epsfig}
\usepackage{times}
\usepackage{amssymb}
\usepackage{amsthm}
\usepackage{setspace}
\usepackage{cite}

\usepackage{algorithmic}  
\usepackage{algorithm}

\usepackage{shadow}
\usepackage{fancybox}
\usepackage{fancyhdr}

\def\w{{\bf w}}

\def\x{{\bf x}}

\def\x{{\mathbf x}}

\def\w{{\bf w}}

\def\x{{\bf x}}

\def\a{{\bf a}}
\def\b{{\bf b}}
\def\c{{\bf c}}

\def\h{{\bf h}}

\def\be{\begin{equation}}
\def\ee{\end{equation}}
\def\ba{\left[\begin{array}}
\def\ea{\end{array}\right]}

\def\w{{\bf w}}

\def\x{{\bf x}}

\def\a{{\bf a}}
\def\b{{\bf b}}
\def\c{{\bf c}}

\def\1{{\bf 1}}

\def\g{{\bf g}}
\def\0{{\bf 0}}

\newtheorem{lemma}{Lemma}

\begin{document}

\begin{singlespace}

\title {Regularly random duality 
}
\author{
\textsc{Mihailo Stojnic}
\\
\\
{School of Industrial Engineering}\\
{Purdue University, West Lafayette, IN 47907} \\
{e-mail: {\tt mstojnic@purdue.edu}} }
\date{}
\maketitle

\centerline{{\bf Abstract}} \vspace*{0.1in}

In this paper we look at a class of random optimization problems. We discuss ways that can help determine typical behavior of their solutions. When the dimensions of the optimization problems are large such an information often can be obtained without actually solving the original problems. Moreover, we also discover that fairly often one can actually determine many quantities of interest (such as, for example, the typical optimal values of the objective functions) completely analytically. We present a few general ideas and emphasize that the range of applications is enormous.

\vspace*{0.25in} \noindent {\bf Index Terms: Linear constraints; duality}.

\end{singlespace}

\section{Introduction}
\label{sec:back}

We start by looking at a class of very simple optimization problems. Namely, we will look at a linearly constrained optimization problems. Such problems can be formulated in the following fairly general way:
\begin{eqnarray}
\min_{\x} & & f(\x)\nonumber \\
\mbox{subject to} & & A\x=0\nonumber \\
& & B\x\leq 0, \label{eq:lincons}
\end{eqnarray}
for concreteness we will assume that $A$ is an $m_1\times n$ matrix from $R^{m_1\times n}$ and $B$ is an $m_2\times n$ matrix from $R^{m_2\times n}$. Also, it is rather clear, but we still mention that $f(\x):R^n\rightarrow R$, is what we will call the objective function. Also if one looks at the problem given in (\ref{eq:lincons}) the first thing that comes to mind is that it is a linearly constrained optimization problem (see, e.g. \cite{BV}). So, there is really nothing specific about it beyond that depending on the type of function $f(\x)$ its objective value could be either bounded or unbounded and the problem can be either feasible or not. To make the exposition easier we will assume that whenever something in our exposition can be such that the objective could be unbounded or even nonexistent then such a scenario is not the subject of our discussion in this paper. Or in other words, we will assume that we look only at the scenarios where the objective values can be computed and are properly bounded. Another alternative would be, if say the problem above is unbounded, to simply add constraints that would insure boundedness of the objective value; or on the other hand, if the problem above is say infeasible, to simply remove some of the constraints until it becomes feasible. We will occasionally throughout the paper look at a few scenarios where we would need to force the boundedness. However, since there will be those where we will ignore it we simply preface it right here before we proceed with further presentation.

Now, going back to the optimization problem given in (\ref{eq:lincons}). Determining the solution of this problem and the optimal value of its objective function is of course the ultimate goal. The type of function $f(\x)$ is typically what determines if this problem can be solved in polynomial time or not. For a moment let us assume that $f(\x)$ is such that (\ref{eq:lincons}) can be solved in polynomial time (in this paper whenever we say polynomial time, we mean it roughly speaking, i.e. without all the details related to what is typically in the complexity theory called strongly polynomial and all other subtleties that come with considerations similar to that). From an algorithmic point of view the above problem is then typically considered as solvable. Our interest in this paper will be slightly different from this classical approach. We will look at a class of these problems and discuss whether or not is it possible to analytically determine the optimal value of the objective function. Of course, if the dimension of the problem, $n$, is small (say $n=2$ or $n=3$) it is highly likely (no matter how complicated $f(\x)$ can be) that (\ref{eq:lincons}) can be solved analytically. As one may guess our interest will not be in such small dimensional scenarios either. Instead we will typically look at large values of $n$ and all other dimensions. Moreover, to facilitate writing, we will typically assume the so-called \emph{linear} regime, i.e. we will assume that all dimensions in this paper are large but linearly proportional to $n$. For example, in (\ref{eq:lincons}) we will assume that $m_1=\alpha_1 n$ and $m_2=\alpha_2 n$ where both, $\alpha_1$ and $\alpha_2$ are constants independent of $n$.

Now, if the dimensions in (\ref{eq:lincons}) are large and our goal is to solve it analytically how exactly do we plan to go about it. Well, there is really not much we would be able to say right away for two reasons: 1) we have not specified $f(\x)$ and dealing with an unspecified $f(\x)$ could be unpredictable and in fact quite often impossible; 2) the dimension of the problem is large which means that the number of constraints is large as well and moreover in a general setup that we assume they all act on all components of $\x$, i.e. on all $\{\x_1,\x_2,\dots,\x_n\}$. While we will not change much in our specifications of $f(\x)$ we will look for a glimmer of hope in a particular type of constraints. In other words we will leave the first of the above reasons aside and try to deal with the second one hoping that that alone will introduce enough simplifications so that eventually even the first reason is not that much of a problem. There are many ways how one deal with sets $A\x=0, B\x\leq 0$. Our approach will be a random one. More specifically, we will assume that the set of constraints is drawn from a probability distribution. Since matrices $A$ and $B$ essentially determine the constraints we will assume that they are the objects that are random. Moreover, to make the presentation easier and to introduce a bit more of concreteness we will also assume that all components of both, $A$ and $B$, are i.i.d standard normal random variables. This effectively establishes problem in (\ref{eq:lincons}) as a random optimization problem and that is the class of the optimization problems that will be the subject of our study in this paper. Fairly often, in the theory of random optimization problems one looks at the objective values that are also random functions of unknown $\x$. Our entire exposition can easily be adapted to encompass such a scenario as well. However, we find it easier from the presentation point of view to assume that $f(\x)$ is actually a deterministic function.

While there is a quite large literature on studying algorithmic aspects of random optimization problems we stop short of reviewing it here. The main reason is that here we are not interested in a specific instance of a certain optimization problem but rather a large class of optimization problems and it would be fairly hard to cover all the relevant work without missing some specific portions of it. We do however mention that the problems we study here are very generic and the literature on any of its particular instances would be a solid subreview. We also mention that our exposition does not rely on using any of the results known for any specific instance. In that sense the reader is not really even required to have pretty much any background in optimization theory beyond a few classical concept that will be rather obvious from our presentation. Moreover, any such concepts will be fairly general and not tailored in any way for the classes of problems we study here.

Now that we have a setup of the introductory problem that we will look at we briefly describe what we will present in the rest of the paper and how the paper  will be organized. In Section \ref{sec:randlincons} we look at problem (\ref{eq:lincons}) in the above mentioned random context and present several observations that can be useful in analytically studying typical probabilistic behavior of the solutions of such problems. In  Section \ref{sec:sophopt} we then study two more general versions of the original problem (\ref{eq:lincons}), namely nonhomogeneous linear constraints and additional functional constraints. In Section \ref{sec:examplehomf} we then look at several particular objective functions and present in details how the mechanisms of Section \ref{sec:sophopt} work. In Section \ref{sec:conc} we give a brief discussion and present several conclusion related to presented results.

\section{Random linearly constrained programs}
\label{sec:randlincons}

In this section we look at problem (\ref{eq:lincons}) in a statistical scenario. As mentioned above, for concreteness we assume that in
\begin{eqnarray}
\xi(f,A,B)=\min_{\x} & & f(\x)\nonumber \\
\mbox{subject to} & & A\x=0\nonumber \\
& & B\x\leq 0, \label{eq:randlincons1}
\end{eqnarray}
all components of matrices $A$ and $B$ are i.i.d. standard normals. Since the assumed scenario is random we also need to revisit one of the assumptions we have made right after (\ref{eq:lincons}). Namely, we stated that we will ignore all situations where the objective function is unbounded. Given the statistical scenario we will slightly modified such a statement by saying that we will assume that the objective in (\ref{eq:randlincons1}) is bounded with overwhelming probability (under overwhelming probability we in this paper assume a probability that is no more than a number exponentially decaying in $n$ away from $1$). Under such an assumption we then proceed with the following transformation of (\ref{eq:randlincons1})
\begin{eqnarray}
\xi(f,A,B)=\min_{\x}\max_{\nu,\lambda} & & f(\x)+\nu^TA\x+\lambda^TB\x\nonumber \\
\mbox{subject to} & & \lambda_i\geq 0,i=1,2,\dots,m_2. \label{eq:randlincons2}
\end{eqnarray}
In the rest of this section we will present a strategy that can be helpful in obtaining a probabilistic view of quantity $\xi(f,A,B)$. We will split the presentation in two parts. In the first part we will present a lower bound type of strategy whereas in the second part we will present an upper bound type of strategy.

\subsection{Lower-bounding strategy}
\label{sec:lbrandlincons}

We will invoke the results of the following lemma which is a slightly modified version of Lemma 3.1 from \cite{Gordon88} (which is a direct consequence of Theorem B from \cite{Gordon88}).
\begin{lemma}
Let $A$ be an $m_1\times n$ matrix with i.i.d. standard normal components and let $B$ be an $m_2\times n$ matrix with i.i.d. standard normal components. Let $\g$ and $\h$ be $n\times 1$ and $(m_1+m_2)\times 1$ vectors, respectively, with i.i.d. standard normal components. Also, let $g$ be a standard normal random variable. Then
\begin{multline}
P(\min_{\x}\max_{\lambda\geq 0,\nu}(\nu^T A\x+\lambda^T B\x +\|\begin{bmatrix}\nu^T \lambda^T\end{bmatrix}\|_2\|\x\|_2 g-\zeta_{\x,\nu,\lambda}^{(l)})\geq 0)\\
\geq P(\min_{\x}\max_{\lambda\geq 0,\nu}(\|\begin{bmatrix}\nu^T \lambda^T\end{bmatrix}\|_2\g^T\x+\|\x\|_2\h^T\begin{bmatrix}\nu^T \lambda^T\end{bmatrix}^T-\zeta_{\x,\nu,\lambda}^{(l)})\geq 0).\label{eq:problemma}
\end{multline}\label{eq:lbgorlemma}
\end{lemma}
\begin{proof}
The proof follows from Theorem B from \cite{Gordon88} after a fairly obvious modification of the proof of Lemma 3.1 given in \cite{Gordon88}.
\end{proof}
Let $\zeta_{\x,\nu,\lambda}^{(l)}=\epsilon_{5}^{(g)}\sqrt{n}\|\begin{bmatrix}\nu^T \lambda^T\end{bmatrix}\|_2\|\x\|_2-f(\x)+\xi_D^{(l)}(f)$ with $\epsilon_{5}^{(g)}>0$ being an arbitrarily small constant independent of $n$ and $\xi_D^{(l)}(f)$ being a fixed number that we will discuss later in great detail. Also, let $\h=[\h_A^T \h_B^T]^T$, where $\h_A$ is the first $m_1$ components of $\h$ and $\h_B$ is the last $m_2$ components of $\h$. We will first look at the right-hand side of the inequality in (\ref{eq:problemma}). The following is then the probability of interest
\begin{equation}
P(\min_{\x}\max_{\lambda\geq 0,\nu}(\|\begin{bmatrix}\nu^T \lambda^T\end{bmatrix}\|_2\g^T\x+\|\x\|_2\h^T\begin{bmatrix}\nu^T \lambda^T\end{bmatrix}^T-\zeta_{\x,\nu,\lambda}^{(l)})\geq 0).\label{eq:probanal0}
\end{equation}
Before further looking at this probability we will look in a bit more detail at the optimization problem inside the probability. We first denote
\begin{equation}
L=\min_{\x}\max_{\lambda\geq 0,\nu}(\|\begin{bmatrix}\nu^T \lambda^T\end{bmatrix}\|_2\g^T\x+\|\x\|_2\h^T\begin{bmatrix}\nu^T \lambda^T\end{bmatrix}^T-\zeta_{\x,\nu,\lambda}^{(l)}).\label{eq:probanal1}
\end{equation}
Replacing the value of $\zeta_{\x,\nu,\lambda}^{(l)}$ we further have
\begin{multline}
L  = \min_{\x}\max_{\lambda\geq 0,\nu}(\|\begin{bmatrix}\nu^T \lambda^T\end{bmatrix}\|_2\g^T\x+\|\x\|_2\h^T\begin{bmatrix}\nu^T \lambda^T\end{bmatrix}^T-\zeta_{\x,\nu,\lambda}^{(l)})\\
     =  \min_{\x}\max_{\lambda\geq 0,\nu}(\|\begin{bmatrix}\nu^T \lambda^T\end{bmatrix}\|_2\g^T\x+\|\x\|_2\h^T\begin{bmatrix}\nu^T \lambda^T\end{bmatrix}^T-\epsilon_{5}^{(g)}\sqrt{n}\|\begin{bmatrix}\nu^T \lambda^T\end{bmatrix}\|_2\|\x\|_2+f(\x)-\xi_D^{(l)}(f).\label{eq:probanal2}
\end{multline}
One can now do the inner maximization for a fixed $\x$ and fixed $\|\begin{bmatrix}\nu^T \lambda^T\end{bmatrix}\|_2$. We then get
\begin{equation}
\hspace{-.5in}L  =  \min_{\x}\max_{\lambda\geq 0,\nu}(\|\begin{bmatrix}\nu^T \lambda^T\end{bmatrix}\|_2\g^T\x+\|\x\|_2\|\begin{bmatrix}\nu^T \lambda^T\end{bmatrix}^T\|_2\sqrt{\|\h_A\|_2^2+\|\h_{B+}\|_2^2}-\epsilon_{5}^{(g)}\sqrt{n}\|\begin{bmatrix}\nu^T \lambda^T\end{bmatrix}\|_2\|\x\|_2+f(\x)-\xi_D^{(l)}(f),\label{eq:probanal3}
\end{equation}
where $\h_{B+}$ is the vector comprised only of non-negative components of $\h_B$. To make sure that $L$ remains bounded we further have
\begin{eqnarray}
L=\min_{\x} & & f(\x)-\xi_D^{(l)}(f)\nonumber \\
\mbox{subject to} & & \g^T\x+\|\x\|_2\|(\sqrt{\|\h_A\|_2^2+\|\h_{B+}\|_2^2}-\epsilon_{5}^{(g)}\sqrt{n})\leq 0. \label{eq:probanal4}
\end{eqnarray}
Combining (\ref{eq:probanal0}), (\ref{eq:probanal1}), and (\ref{eq:probanal4}) one then has for the left hand side of (\ref{eq:problemma})
\begin{equation}
P(\min_{\x}\max_{\lambda\geq 0,\nu}(\|\begin{bmatrix}\nu^T \lambda^T\end{bmatrix}\|_2\g^T\x+\|\x\|_2\h^T\begin{bmatrix}\nu^T \lambda^T\end{bmatrix}^T-\zeta_{\x,\nu,\lambda}^{(l)})\geq 0)=P(L\geq 0),\label{eq:probanal5}
\end{equation}
with $L$ given in (\ref{eq:probanal4}). Since $\h_A$ is a vector of $m_1$ i.i.d. standard normal variables and $\h_B$ is a vector of $m_2$ i.i.d. standard normal variables it is rather trivial that
\begin{equation*}
P(\sqrt{\|\h_A\|_2^2+\|\h_{B+}\|_2^2}>(1-\epsilon_{1}^{(m)})\sqrt{m_1+m_2/2})\geq 1-e^{-\epsilon_{2}^{(m)} (m_1+m_2/2)},
\end{equation*}
where $\epsilon_{1}^{(m)}>0$ is an arbitrarily small constant and $\epsilon_{2}^{(m)}$ is a constant dependent on $\epsilon_{1}^{(m)}$ but independent of $n$. Then one can modify (\ref{eq:probanal4}) and (\ref{eq:probanal5}) in the following way
\begin{multline}
P(\min_{\x}\max_{\lambda\geq 0,\nu}(\|\begin{bmatrix}\nu^T \lambda^T\end{bmatrix}\|_2\g^T\x+\|\x\|_2\h^T\begin{bmatrix}\nu^T \lambda^T\end{bmatrix}^T-\zeta_{\x,\nu,\lambda}^{(l)})\geq 0)=P(L\geq 0)\\\geq (1-e^{-\epsilon_{2}^{(m)} (m_1+m_2/2)})P(L^{(1)}\geq 0),\label{eq:probanal6}
\end{multline}
where $L^{(1)}$ is
\begin{eqnarray}
L^{(1)}=\min_{\x} & & f(\x)-\xi_D^{(l)}(f)\nonumber \\
\mbox{subject to} & & \g^T\x+\|\x\|_2((1-\epsilon_{1}^{(m)})\sqrt{m_1+m_2/2}-\epsilon_{5}^{(g)}\sqrt{n})\leq 0. \label{eq:probanal7}
\end{eqnarray}

We now look at the left-hand side of the inequality in (\ref{eq:problemma}).
\begin{multline}
P(\min_{\x}\max_{\lambda\geq 0,\nu}(\nu^T A\x+\lambda^T B\x +\|\begin{bmatrix}\nu^T \lambda^T\end{bmatrix}\|_2\|\x\|_2 g-\zeta_{\x,\nu,\lambda}^{(l)})\geq 0)\\=
P(\min_{\x}\max_{\lambda\geq 0,\nu}(\nu^T A\x+\lambda^T B\x+f(\w)-\xi_D^{(l)}(f)+\|\begin{bmatrix}\nu^T \lambda^T\end{bmatrix}\|_2\|\x\|_2(g-\epsilon_{5}^{(g)}\sqrt{n}))\geq 0).\label{eq:leftprobanal0}
\end{multline}
Since $P(g\leq\epsilon_{5}^{(g)}\sqrt{n})> 1-e^{-\epsilon_{6}^{(g)} n}$ (where $\epsilon_{6}^{(g)}$ is, as all other $\epsilon$'s in this paper are, independent of $n$) from (\ref{eq:leftprobanal0}) we have
\begin{multline}
P(\min_{\x}\max_{\lambda\geq 0,\nu}(\nu^T A\x+\lambda^T B\x +\|\begin{bmatrix}\nu^T \lambda^T\end{bmatrix}\|_2\|\x\|_2 g-\zeta_{\x,\nu,\lambda}^{(l)})\geq 0)\\
\leq
(1-e^{-\epsilon_{6}^{(g)} n})P(\min_{\x}\max_{\lambda\geq 0,\nu}(\nu^T A\x+\lambda^T B\x+f(\w)-\xi_D^{(l)}(f))\geq 0)+e^{-\epsilon_{6}^{(g)} n}.\label{eq:leftprobanal1}
\end{multline}
Connecting (\ref{eq:problemma}), (\ref{eq:probanal5}), (\ref{eq:probanal6}), and (\ref{eq:leftprobanal1}) we obtain
\begin{multline}
P(\min_{\x}\max_{\lambda\geq 0,\nu}(\nu^T A\x+\lambda^T B\x+f(\w)-\xi_D^{(l)}(f))\geq 0)
\geq\\
\frac{(1-e^{-\epsilon_{2}^{(m)} (m_1+m_2/2)})}{(1-e^{-\epsilon_{6}^{(g)} n})}
P(L^{(1)}\geq 0)
-\frac{e^{-\epsilon_{6}^{(g)} n}}{(1-e^{-\epsilon_{6}^{(g)} n})},\label{eq:leftprobanal2}
\end{multline}
where $L^{(1)}$ is as given in (\ref{eq:probanal7}). A further combination of (\ref{eq:randlincons2}) and (\ref{eq:leftprobanal2}) gives
\begin{equation}
P(\xi(f,A,B)-\xi_D^{(l)}(f)\geq 0)
\geq
\frac{(1-e^{-\epsilon_{2}^{(m)} m})}{(1-e^{-\epsilon_{6}^{(g)} n})}
P(L^{(1)}\geq 0)
-\frac{e^{-\epsilon_{6}^{(g)} n}}{(1-e^{-\epsilon_{6}^{(g)} n})}.\label{eq:leftprobanal3}
\end{equation}

We are now in position to state the following lemma which is the first of results that we will present that relates to the optimal value of the objective of (\ref{eq:randlincons1}).

\begin{lemma}(Lower bound)
\label{thm:reversemesh} Let $A$ be an $m_1\times n$ matrix with i.i.d. standard normal components. Let $B$ be an $m_2\times n$ matrix with i.i.d. standard normal components. Assume that $n$ is large and that $m_1=\alpha_1 n$ and $m_2=\alpha_2 n$ where $\alpha_1$ and $\alpha_2$ are constants independent of $n$. Let $f(\x):R^n\rightarrow R$ be a given function and let $\xi(f,A,B)$ be the objective value of the optimization problem in (\ref{eq:randlincons1}). Assume that $f(\x)$ is such that $|\xi(f,A,B)|<\infty$ with overwhelming probability. Further let $\g$ be an $n\times 1$ vector with i.i.d. standard normal components. Let $\epsilon$'s in (\ref{eq:probanal7}) be arbitrarily small constants and let $\xi_D^{(l)}(f)$ be the largest scalar so that $L^{(1)}$ defined in (\ref{eq:probanal7}) is non-negative with overwhelming probability. Then,
\begin{equation*}
\lim_{n\rightarrow\infty}P(\xi(f,A,B)>\xi_D^{(l)}(f))=1.
\end{equation*}
\label{lemma:lblincons}
\end{lemma}

\begin{proof}
Follows from the previous discussion.
\end{proof}

While the above lemma may sound a bit dry it is often a fairly powerful tool to deal with random linearly constrained programs. Its power essentially lies in potential simplicity of the auxiliary optimization program (\ref{eq:probanal7}). It is relatively easy to see that the optimization problem in (\ref{eq:probanal7}) is substantially simpler than the original one given in (\ref{eq:randlincons1}). Still, there is no guarantee that (\ref{eq:probanal7}) is always solvable. That would certainly depend on the structure of function $f(\x)$. Also, not only that (\ref{eq:probanal7}) needs to be solvable, one should also be able to show that its solution behaves ``nicely", i.e. one should be able to find a quantity $\xi_D^{(l)}$ that is almost certain to be smaller than the optimal value of the objective of (\ref{eq:probanal7}). We will towards the end of the paper demonstrate how the results of this lemma can be used in practice on a small example. The key in such an example (as well as in any example where the above lemma is to be of any use) will be ability to probabilistically handle much simpler program (\ref{eq:probanal7}). Before proceeding with further generic considerations of (\ref{eq:randlincons1}) we will in the next subsection we present a corresponding upper-bounding strategy for a probabilistic characterization of the objective in (\ref{eq:randlincons1}).

\subsection{Virtual upper-bounding strategy}
\label{sec:ubrandlincons}

Before we proceed with the detail presentation we should make more explicit the following point. Namely, what we presented in the previous subsection is a concept that is mathematically speaking always correct, i.e. as long as the problem in (\ref{eq:randlincons1}) is deterministically solvable and its solution probabilistically speaking bounded. Now, while the concept is correct it is just a lower bound type of approach. Moreover, the concept is correct and it relies on a potential simplicity of (\ref{eq:probanal7}). So it will be useful if (\ref{eq:probanal7}) can be handled. However, no matter if (\ref{eq:probanal7}) can be handled or not, the entire concept remains valid with very minimal assumptions on $f(\x)$ (in fact, assumption that $f(\x)$ is such that (\ref{eq:randlincons1}) is bounded seems as pretty much unavoidable as long as solving (\ref{eq:randlincons1}) is to have any reasonable practical sense). On the other hand the strategy that we will present below will not work generically, i.e. it will require additional assumptions on $f(\x)$ beyond those mentioned in the previous subsection. Since these assumptions may or may not hold we will preface our presentation by saying that the upper-bounding strategy that we show below is in a way a virtual strategy. Of course, we should add that the strategy is not purely virtual. Quite contrary, it fairly often works; in fact, roughly speaking, it works almost exactly as the complexity theory works, i.e. it is fairly similar to the following paradigm ``as long as (\ref{eq:randlincons1}) is computationally doable in a reasonable amount of time the strategy will be working well". Of course, this is a fairly informal statement without any mathematically rigorous type of language. To establish the above statement on a more mathematically rigorous level requires a presentation that goes way beyond the scope of this paper and will pursue it elsewhere. We do mention that such a presentation does not contain almost any further conceptual insight, i.e. the core of the ideas is already here. However, it does require an enormous amount of mathematical detailing which we choose to skip to avoid ruining the elegance of the presentation that we attempt to achieve here.

Going back to (\ref{eq:randlincons1}), in this section we will essentially attempt to mimic the presentation of the previous subsection. To that end we start by recalling that our object of interest is the following linearly constrained optimization problem:
\begin{eqnarray}
\xi(f,A,B)=\min_{\x} & & f(\x)\nonumber \\
\mbox{subject to} & & A\x=0\nonumber \\
& & B\x\leq 0, \label{eq:ubrandlincons1}
\end{eqnarray}
which after a bit of juggling becomes
\begin{equation}
\xi(f,A,B)=\min_{\x}\max_{\lambda\geq 0,\nu} f(\x)+\nu^TA\x+\lambda^TB\x. \label{eq:ubrandlincons2}
\end{equation}
Now, we recall that $A$ and $B$ are random matrices and the above optimization problems are random. Given their randomness sometimes they can be solvable sometimes they may not be solvable. They may be unsolvable due to the fact that they are not feasible or that they are feasible but the value of the objective function is unbounded. However, as we did in the previous subsections, we leave all these unfavorable scenarios aside and preface our presentation assuming that
$|\xi(f,A,B)|\leq \infty$ with overwhelming probability.

A this point we will attempt to transform (\ref{eq:ubrandlincons2}) assuming that $f(\x)$ is such that the transformation is mathematically possible. Namely, let $f(\x)$ be such that
\begin{eqnarray}
\xi(f,A,B) & = & \min_{\x}\max_{\lambda\geq 0,\nu} f(\x)+\nu^TA\x+\lambda^TB\x\nonumber \\
& = & \max_{\lambda\geq 0,\nu}\min_{\x} f(\x)+\nu^TA\x+\lambda^TB\x. \label{eq:ubrandlincons3}
\end{eqnarray}
Assuming that (\ref{eq:ubrandlincons3}) holds one can then further write
\begin{eqnarray}
\xi(f,A,B) & = & \max_{\lambda\geq 0,\nu}\min_{\x} f(\x)+\nu^TA\x+\lambda^TB\x\nonumber \\
& = & -\min_{\lambda\geq 0,\nu}\max_{\x} -f(\x)-\nu^TA\x-\lambda^TB\x \label{eq:ubrandlincons4}
\end{eqnarray}
and
\begin{equation}
-\xi(f,A,B) = \min_{\lambda\geq 0,\nu}\max_{\x} -f(\x)-\nu^TA\x-\lambda^TB\x. \label{eq:ubrandlincons5}
\end{equation}
Similarly to what was done in the previous subsection we will utilize the results of the following lemma which is a slightly modified version of Lemma 3.1 from \cite{Gordon88} and an upper-bounding analogue to lower-bounding Lemma \ref{eq:lbgorlemma}.
\begin{lemma}
Let $A$ be an $m_1\times n$ matrix with i.i.d. standard normal components and let $B$ be an $m_2\times n$ matrix with i.i.d. standard normal components. Let $\g$ and $\h$ be $n\times 1$ and $(m_1+m_2)\times 1$ vectors, respectively, with i.i.d. standard normal components. Also, let $g$ be a standard normal random variable. Then
\begin{multline}
P(\min_{\lambda\geq 0,\nu}\max_{\x}(-\nu^T A\x-\lambda^T B\x +\|\begin{bmatrix}\nu^T \lambda^T\end{bmatrix}\|_2\|\x\|_2 g-\zeta_{\x,\nu,\lambda}^{(u)})\geq 0)\\
\geq P(\min_{\lambda\geq 0,\nu}\max_{\x}(\|\begin{bmatrix}\nu^T \lambda^T\end{bmatrix}\|_2\g^T\x+\|\x\|_2\h^T\begin{bmatrix}\nu^T \lambda^T\end{bmatrix}^T-\zeta_{\x,\nu,\lambda}^{(u)})\geq 0).\label{eq:ubproblemma}
\end{multline}\label{eq:ubgorlemma}
\end{lemma}
\begin{proof}
The proof follows from Theorem B from \cite{Gordon88} after a fairly obvious modification of the proof of Lemma 3.1 given in \cite{Gordon88}.
\end{proof}
Let $\zeta_{\x,\nu,\lambda}^{(u)}=\epsilon_{5}^{(g)}\sqrt{n}\|\begin{bmatrix}\nu^T \lambda^T\end{bmatrix}\|_2\|\x\|_2+f(\x)-\xi_D^{(u)}(f)$ with $\epsilon_{5}^{(g)}>0$ being an arbitrarily small constant independent of $n$ and $\xi_D^{(u)}(f)$ being a fixed number that we will discuss later in great detail. As in the previous subsection, let $\h=[\h_A^T \h_B^T]^T$, where $\h_A$ is the first $m_1$ components of $\h$ and $\h_B$ is the last $m_2$ components of $\h$. As in the previous subsection we will first look at the right-hand side of the inequality in (\ref{eq:ubproblemma}). The following is then the probability of interest
\begin{equation}
P(\min_{\lambda\geq 0,\nu}\max_{\x}(\|\begin{bmatrix}\nu^T \lambda^T\end{bmatrix}\|_2\g^T\x+\|\x\|_2\h^T\begin{bmatrix}\nu^T \lambda^T\end{bmatrix}^T-\zeta_{\x,\nu,\lambda}^{(u)})\geq 0).\label{eq:ubprobanal0}
\end{equation}
Before looking further at this probability we will look in a bit more detail at the optimization problem inside the probability. We first denote
\begin{equation}
U=\min_{\lambda\geq 0,\nu}\max_{\x}(\|\begin{bmatrix}\nu^T \lambda^T\end{bmatrix}\|_2\g^T\x+\|\x\|_2\h^T\begin{bmatrix}\nu^T \lambda^T\end{bmatrix}^T-\zeta_{\x,\nu,\lambda}^{(u)}).\label{eq:ubprobanal1}
\end{equation}
Replacing the value of $\zeta_{\x,\nu,\lambda}^{(u)}$ we further have
\begin{multline}
U  = \min_{\lambda\geq 0,\nu}\max_{\x}(\|\begin{bmatrix}\nu^T \lambda^T\end{bmatrix}\|_2\g^T\x+\|\x\|_2\h^T\begin{bmatrix}\nu^T \lambda^T\end{bmatrix}^T-\zeta_{\x,\nu,\lambda}^{(u)})\\
     =  \min_{\lambda\geq 0,\nu}\max_{\x}(\|\begin{bmatrix}\nu^T \lambda^T\end{bmatrix}\|_2\g^T\x+\|\x\|_2\h^T\begin{bmatrix}\nu^T \lambda^T\end{bmatrix}^T-\epsilon_{5}^{(g)}\sqrt{n}\|\begin{bmatrix}\nu^T \lambda^T\end{bmatrix}\|_2\|\x\|_2-f(\x)+\xi_D^{(u)}(f)).\label{eq:ubprobanal2}
\end{multline}
From (\ref{eq:ubprobanal2}) one then has
\begin{multline}
U  =  \min_{\lambda\geq 0,\nu}\max_{\x}(\|\begin{bmatrix}\nu^T \lambda^T\end{bmatrix}\|_2\g^T\x+\|\x\|_2\h^T\begin{bmatrix}\nu^T \lambda^T\end{bmatrix}^T-\epsilon_{5}^{(g)}\sqrt{n}\|\begin{bmatrix}\nu^T \lambda^T\end{bmatrix}\|_2\|\x\|_2-f(\x)+\xi_D^{(u)}(f))\\
\geq \max_{\x}\min_{\lambda\geq 0,\nu}(\|\begin{bmatrix}\nu^T \lambda^T\end{bmatrix}\|_2\g^T\x+\|\x\|_2\h^T\begin{bmatrix}\nu^T \lambda^T\end{bmatrix}^T-\epsilon_{5}^{(g)}\sqrt{n}\|\begin{bmatrix}\nu^T \lambda^T\end{bmatrix}\|_2\|\x\|_2-f(\x)+\xi_D^{(u)}(f))\\
 = - \min_{\x}\max_{\lambda\geq 0,\nu}(-\|\begin{bmatrix}\nu^T \lambda^T\end{bmatrix}\|_2\g^T\x-\|\x\|_2\h^T\begin{bmatrix}\nu^T \lambda^T\end{bmatrix}^T+\epsilon_{5}^{(g)}\sqrt{n}\|\begin{bmatrix}\nu^T \lambda^T\end{bmatrix}\|_2\|\x\|_2+f(\x)-\xi_D^{(u)}(f)).\label{eq:ubprobanal20}
\end{multline}
One can now do the inner maximization for a fixed $\x$ and fixed $\|\begin{bmatrix}\nu^T \lambda^T\end{bmatrix}\|_2$ to get
\begin{equation}
\hspace{-.5in}U  \geq - \min_{\x}\max_{\lambda\geq 0,\nu}(-\|\begin{bmatrix}\nu^T \lambda^T\end{bmatrix}\|_2\g^T\x+\|\x\|_2\|\begin{bmatrix}\nu^T \lambda^T\end{bmatrix}^T\|_2\sqrt{\|\h_A\|_2^2+\|\h_{B+}\|_2^2}+\epsilon_{5}^{(g)}\sqrt{n}\|\begin{bmatrix}\nu^T \lambda^T\end{bmatrix}\|_2\|\x\|_2+f(\x)-\xi_D^{(u)}(f)),\label{eq:ubprobanal3}
\end{equation}
where as in the previous subsection $\h_{B+}$ is vector comprise of only non-negative components of $\h_B$. To make sure that the quantity on the right-hand side remains bounded we further have
\begin{eqnarray}
U\geq-\min_{\x} & & f(\x)-\xi_D^{(l)}(f)\nonumber \\
\mbox{subject to} & & -\g^T\x+\|\x\|_2\|(\sqrt{\|\h\|_2^2+\|\h_{B+}\|_2^2}+\epsilon_{5}^{(g)}\sqrt{n})\leq 0. \label{eq:ubprobanal4}
\end{eqnarray}
Let
\begin{eqnarray}
U^{(0)}=\min_{\x} & & f(\x)-\xi_D^{(l)}(f)\nonumber \\
\mbox{subject to} & & -\g^T\x+\|\x\|_2\|(\sqrt{\|\h\|_2^2+\|\h_{B+}\|_2^2}+\epsilon_{5}^{(g)}\sqrt{n})\leq 0. \label{eq:ubprobanal5}
\end{eqnarray}
Combining (\ref{eq:ubprobanal0}), (\ref{eq:ubprobanal1}), and (\ref{eq:ubprobanal4}) one then has for the left hand side of (\ref{eq:ubproblemma})
\begin{equation}
P(\min_{\x}\max_{\lambda\geq 0,\nu}(\|\begin{bmatrix}\nu^T \lambda^T\end{bmatrix}\|_2\g^T\x+\|\x\|_2\h^T\begin{bmatrix}\nu^T \lambda^T\end{bmatrix}^T-\zeta_{\x,\nu,\lambda}^{(u)})\geq 0)=P(U\geq 0)\geq P(-U^{(0)}\geq 0)=P(U^{(0)}\leq 0),\label{eq:ubprobanal6}
\end{equation}
with $U^{(0)}$ as given in (\ref{eq:ubprobanal5}). Since $\h_A$ is a vector of $m_1$ i.i.d. standard normal variables and $\h_B$ is a vector of $m_2$ i.i.d. standard normal variables it is rather trivial that
\begin{equation*}
P(\sqrt{\|\h\|_2^2+\|\h_{B+}\|_2^2}<(1+\epsilon_{1}^{(m)})\sqrt{m_1+m_2/2})\geq 1-e^{-\epsilon_{2}^{(m)} (m_1+m_2/2)},
\end{equation*}
where we recall that as in the previous subsection $\epsilon_{1}^{(m)}>0$ is an arbitrarily small constant and $\epsilon_{2}^{(m)}$ is a constant dependent on $\epsilon_{1}^{(m)}$ but independent of $n$. Then one can modify (\ref{eq:ubprobanal5}) and (\ref{eq:ubprobanal6}) in the following way
\begin{multline}
P(\min_{\x}\max_{\lambda\geq 0,\nu}(\|\begin{bmatrix}\nu^T \lambda^T\end{bmatrix}\|_2\g^T\x+\|\x\|_2\h^T\begin{bmatrix}\nu^T \lambda^T\end{bmatrix}^T-\zeta_{\x,\nu,\lambda}^{(u)})\geq 0)\geq P(U^{(0)}\leq 0)\\\geq (1-e^{-\epsilon_{2}^{(m)} (m_1+m_2/2)})P(U^{(1)}\leq 0),\label{eq:ubprobanal7}
\end{multline}
where $U^{(1)}$ is
\begin{eqnarray}
U^{(1)}=\min_{\x} & & f(\x)-\xi_D^{(u)}(f)\nonumber \\
\mbox{subject to} & & -\g^T\x+\|\x\|_2((1+\epsilon_{1}^{(m)})\sqrt{m_1+m_2/2}+\epsilon_{5}^{(g)}\sqrt{n})\leq 0. \label{eq:ubprobanal8}
\end{eqnarray}

We now look at the left-hand side of the inequality in (\ref{eq:ubproblemma}). Essentially we will just need to repeat the corresponding arguments from the previous subsection. A few notational modifications will be in place though. We start with
\begin{multline}
P(\min_{\lambda\geq 0,\nu}\max_{\x}(-\nu^T A\x-\lambda^T B\x +\|\begin{bmatrix}\nu^T \lambda^T\end{bmatrix}\|_2\|\x\|_2 g-\zeta_{\x,\nu,\lambda}^{(u)})\geq 0)\\=
P(\min_{\lambda\geq 0,\nu}\max_{\x}(-\nu^T A\x-\lambda^T B\x-f(\w)+\xi_D^{(u)}(f)+\|\begin{bmatrix}\nu^T \lambda^T\end{bmatrix}\|_2\|\x\|_2(g-\epsilon_{5}^{(g)}\sqrt{n}))\geq 0).\label{eq:ubleftprobanal0}
\end{multline}
Since $P(g\leq\epsilon_{5}^{(g)}\sqrt{n})> 1-e^{-\epsilon_{6}^{(g)} n}$ (where $\epsilon_{6}^{(g)}$ is, as all other $\epsilon$'s in this paper are, independent of $n$) from (\ref{eq:ubleftprobanal0}) we have
\begin{multline}
P(\min_{\lambda\geq 0,\nu}\max_{\x}(-\nu^T A\x-\lambda^T B\x +\|\begin{bmatrix}\nu^T \lambda^T\end{bmatrix}\|_2\|\x\|_2 g-\zeta_{\x,\nu,\lambda}^{(u)})\geq 0)\\
\leq
(1-e^{-\epsilon_{6}^{(g)} n})P(\min_{\lambda\geq 0,\nu}\max_{\x}(-\nu^T A\x-\lambda^T B\x-f(\w)+\xi_D^{(u)}(f))\geq 0)+e^{-\epsilon_{6}^{(g)} n}.\label{eq:ubleftprobanal1}
\end{multline}
Connecting (\ref{eq:ubproblemma}), (\ref{eq:ubprobanal6}), (\ref{eq:ubprobanal7}), and (\ref{eq:ubleftprobanal1}) we obtain
\begin{multline}
P(\min_{\lambda\geq 0,\nu}\max_{\x}(-\nu^T A\x-\lambda^T B\x-f(\w)+\xi_D^{(u)}(f))\geq 0)
\geq\\
\frac{(1-e^{-\epsilon_{2}^{(m)} (m_1+m_2/2)})}{(1-e^{-\epsilon_{6}^{(g)} n})}
P(U^{(1)}\leq 0)
-\frac{e^{-\epsilon_{6}^{(g)} n}}{(1-e^{-\epsilon_{6}^{(g)} n})},\label{eq:ubleftprobanal2}
\end{multline}
where $U^{(1)}$ is as given in (\ref{eq:ubprobanal8}). A further combination of (\ref{eq:ubrandlincons2}) and (\ref{eq:ubleftprobanal2}) gives
\begin{equation}
P(-\xi(f,A,B)+\xi_D^{(u)}(f)> 0)
\geq
\frac{(1-e^{-\epsilon_{2}^{(m)} m})}{(1-e^{-\epsilon_{6}^{(g)} n})}
P(U^{(1)}\leq 0)
-\frac{e^{-\epsilon_{6}^{(g)} n}}{(1-e^{-\epsilon_{6}^{(g)} n})}.\label{eq:ubleftprobanal3}
\end{equation}

We are now in position to state the following lemma which is a result that helps create an upper-bound on the optimal value of the objective of (\ref{eq:randlincons1}).

\begin{lemma}(Virtual upper bound)
\label{thm:reversemesh} Let $A$ be an $m_1\times n$ matrix with i.i.d. standard normal components. Let $B$ be an $m_2\times n$ matrix with i.i.d. standard normal components. Assume that $n$ is large and that $m_1=\alpha_1 n$ and $m_2=\alpha_2 n$ where $\alpha_1$ and $\alpha_2$ are constants independent of $n$. Let $f(\x):R^n\rightarrow R$ be a given function and let $\xi(f,A,B)$ be the objective value of the optimization problem in (\ref{eq:randlincons1}). Assume that $f(\x)$ is such that $|\xi(f,A,B)|<\infty$ with overwhelming probability and that (\ref{eq:ubrandlincons3}) holds. Further let $\g$ be an $n\times 1$ vector with i.i.d. standard normal components. Let $\epsilon$'s in (\ref{eq:ubprobanal8}) be arbitrarily small constants and let $\xi_D^{(u)}(f)$ be the smallest scalar so that $U^{(1)}$ defined in (\ref{eq:ubprobanal8}) is non-positive with overwhelming probability. Then,
\begin{equation*}
\lim_{n\rightarrow\infty}P(\xi(f,A,B)<\xi_D^{(u)}(f))=1.
\end{equation*}
\label{lemma:ublincons}
\end{lemma}

\begin{proof}
Follows from the previous discussion.
\end{proof}

As was the case with Lemma \ref{lemma:lblincons}, Lemma \ref{lemma:ublincons} may also sound a bit dry. However, as we mentioned right after Lemma \ref{lemma:lblincons}, Lemma \ref{lemma:ublincons} often turns out to be a fairly powerful tool to deal with random linearly constrained programs. Its major power essentially lies in potential simplicity of the auxiliary optimization program (\ref{eq:ubprobanal8}) (of course excluding a couple of technical details this program is for all practical purposes the same as the one given in (\ref{eq:probanal7})). On the other hand one should keep in mind that Lemma \ref{lemma:ublincons} is a bit more restrictive in that it also requires that $f(\x)$ is such that (\ref{eq:ubrandlincons3}) holds. If one for a moment leaves aside this restriction then the power of the above lemma pretty much relies on one's ability to determine a quantity $\xi_D^{(u)}$ that is almost certain to be larger than the optimal value of $f(\x)$ in (\ref{eq:ubprobanal8}). Of course the smaller $\xi_D^{(u)}$ the better the bound. In a more informal language though, if a duality in (\ref{eq:ubrandlincons3}) holds and if everything else (probabilistically speaking) behaves ``nicely" the success of the above introduced mechanism relies on one's ability to provide a precise probabilistic analysis of (\ref{eq:probanal7}) or (\ref{eq:ubprobanal8}). That is typically highly likely to be possible given that the
optimization program (\ref{eq:probanal7}) (or (\ref{eq:ubprobanal8})) has only one random linear constraint.

\section{More sophisticated optimization programs}
\label{sec:sophopt}

What we presented in the previous section is an often very powerful mechanism to handle linearly constrained optimization programs. One then naturally may wonder is there a way to extend the above results to more general classes of optimization problems. The answer is yes, but in our experience such extensions are typically problem specific. That is of course one of the reasons why we presented the main concepts on a very simple optimization problem. Instead of listing various other types of problems where the mechanism presented here can be used equally successfully we below choose to discuss a few small modifications which will hopefully provide a hint as to how relatively easily the whole framework can be massaged to fit into various other scenarios. All these modifications could have been already included in our original setup. However, we thought that they would make the original problem unnecessary cumbersome and in order to preserve the lightness of the exposition we chose to start with the simplest possible example and then build from there.

\subsection{Non-homogeneous linear constraints}
\label{sec:nonhomcons}

Looking back at problem (\ref{eq:lincons}) one can notice that we started with a set of constraints that is basically homogeneous, i.e. pretty much scaling invariant. In other words for any $\x$ that is feasible in (\ref{eq:lincons}) $c\x$ is feasible as well as long as $c\geq 0$. Typically linear constraints are not necessarily homogeneous and if they are not one has the following more general version of (\ref{eq:lincons})
\begin{eqnarray}
\min_{\x} & & f(\x)\nonumber \\
\mbox{subject to} & & A\x=\a\nonumber \\
& & B\x\leq \b, \label{eq:nonhomlincons1}
\end{eqnarray}
where $\a$ is an $m_1\times 1$ vector from $R^{m_1}$ and analogously $\b$ is an $m_2\times 1$ vector from $R^{m_2}$. Now, given that in this paper we are dealing with random programs, it is natural to wonder if $\a$ and/or $\b$ are random or deterministic (fixed). We will below just sketch how our results easily adapt if $\a$ and $\b$ are deterministic. Essentially, one can pretty much repeat the entire derivation from the previous section. Namely, one can start by defining the optimal value of the objective in (\ref{eq:nonhomlincons1}) as
\begin{eqnarray}
\xi_{nh}(f,A,B)=\min_{\x} & & f(\x)\nonumber \\
\mbox{subject to} & & A\x=\a\nonumber \\
& & B\x\leq \b, \label{eq:nonhomlincons2}
\end{eqnarray}
and write an analogue to (\ref{eq:randlincons2})
\begin{equation}
\xi_{nh}(f,A,B)=\min_{\x}\max_{\lambda\geq 0,\nu} f(\x)+\nu^T A\x+\lambda^T B\x+\nu^T\a+\lambda^T\b. \label{eq:nonhomlincons3}
\end{equation}
One can then repeat the entire definition from the previous section with very minimal and fairly obvious modifications. We skip such an exercise but mention only the critical differences and final results. The only difference in the entire derivation will be the form of the auxiliary programs (\ref{eq:probanal7}) (or (\ref{eq:ubprobanal8})). Since (\ref{eq:probanal7}) (and (\ref{eq:ubprobanal8})) are a more refined version of (\ref{eq:probanal4}) (and (\ref{eq:ubprobanal5})) what will actually change is the structure of these programs. So instead of them one would have
\begin{eqnarray}
L_{nh}^{(0)}=\min_{\x} & & f(\x)-\xi_D^{(l)}(f)\nonumber \\
\mbox{subject to} & & \g^T\x+\sqrt{\|\|\x\|_2\h_A+\a\|_2^2+\|(\|\x\|_2\h_B+\b)_+\|_2^2}-\|\x\|_2\epsilon_{5}^{(g)}\sqrt{n}\leq 0, \label{eq:nonhomprobanal7}
\end{eqnarray}
where $(\|\x\|_2\h_B+\b)_+$ is a vector comprised of non-negative components of vector $\|\x\|_2\h_B+\b$. On the other hand one would have for a corresponding replacement of  (\ref{eq:ubprobanal5})
\begin{eqnarray}
U_{nh}^{(0)}=\min_{\x} & & f(\x)-\xi_D^{(u)}(f)\nonumber \\
\mbox{subject to} & & -\g^T\x+\sqrt{\|\|\x\|_2\h_A+\a\|_2^2+\|(\|\x\|_2\h_B+\b)_+\|_2^2}+\|\x\|_2\epsilon_{5}^{(g)}\sqrt{n}\leq 0. \label{eq:nonhomprobanal8}
\end{eqnarray}
Of course, for all practical purposes programs (\ref{eq:nonhomprobanal7}) and (\ref{eq:nonhomprobanal8}) are basically equivalent. Statement of Lemma \ref{lemma:lblincons} would then remain in place with the only difference being that $L^{(1)}$ should be replaced by $L_{nh}^{(0)}$. Similarly, Lemma \ref{lemma:ublincons} would remain correct with $U^{(1)}$ being replaced by $U_{nh}^{(0)}$ and with an $f(\x)$ being such that the following modified version of (\ref{eq:ubrandlincons3}) holds
\begin{eqnarray}
\xi_{nh}(f,A,B) & = & \min_{\x}\max_{\lambda\geq 0,\nu} f(\x)+\nu^TA\x+\lambda^TB\x+\nu^T\a+\lambda^T\b\nonumber \\
& = & \max_{\lambda\geq 0,\nu}\min_{\x} f(\x)+\nu^TA\x+\lambda^TB\x+\nu^T\a+\lambda^T\b. \label{eq:nonhomubrandlincons3}
\end{eqnarray}
What we presented above is a generic scenario that would work for any given $\a$ and $\b$. Even when $\a$ and $\b$ are generic, one can of course massage it further and remove the randomness of $\h$ as in the definitions of $L^{(1)}$ and $U^{(1)}$ (when $\a$ and $\b$ are random this is even easier). We skip these easy exercises.

\subsection{Additional functional constraints}
\label{sec:addcons}

What we discussed above is an upgrade in the existing set of constraints. Instead one may wonder how mechanism would fare if the linear structure of constraints would be changed to include more general constraints. For example instead of (\ref{eq:lincons}) one may look at its a more general version
\begin{eqnarray}
\min_{\x} & & f(\x)\nonumber \\
\mbox{subject to} & & A\x=0\nonumber \\
& & B\x\leq 0\nonumber\\
& & f_i(\x)\leq 0,i=1,2,\dots,l \label{eq:addmlincons1}
\end{eqnarray}
where each $f_i(\x):R^n\rightarrow R$ is a non-necessarily linear function of $\x$ (of course, there is really no need to restrict on scalar functions; i.e. all the major steps that we present below can be repeated/extended to pretty much any kind of function). Similarly to what we discussed in the previous subsection, these functions can be random of deterministic. To make writing easier we will assume that they are generic, i.e. deterministic. One can then again proceed as above by introducing
\begin{eqnarray}
\xi_{afc}(f,f_1,f_2,\dots,f_l,A,B)=\min_{\x} & & f(\x)\nonumber \\
\mbox{subject to} & & A\x=0\nonumber \\
& & B\x\leq 0\nonumber \\
& & f_i(\x)\leq 0,i=1,2,\dots,l \label{eq:addlincons2}
\end{eqnarray}
and writing an analogue to (\ref{eq:randlincons2})
\begin{equation}
\xi_{afc}(f,f_1,f_2,\dots,f_l,A,B)=\min_{\x}\max_{\lambda\geq 0,\gamma_i\geq 0,\nu} f(\x)+\nu^T A\x+\lambda^T B\x+\sum_{i=1}^{l}\gamma_i f_i(\x). \label{eq:addlincons3}
\end{equation}
One can again then repeat the entire derivation from the previous section with very minimal modifications. As in the previous subsection, we skip such an exercise and only mention the critical differences and final results. As was the case above when we discussed the non-homogeneous linear constraints, the only difference in the repeated derivation will be the form of the auxiliary programs (\ref{eq:probanal7}) (or (\ref{eq:ubprobanal8})). So instead of them one would have
\begin{eqnarray}
L_{afc}^{(1)}=\min_{\x} & & f(\x)-\xi_D^{(l)}(f)\nonumber \\
\mbox{subject to} & & \g^T\x+\|\x\|_2((1-\epsilon_{1}^{(m)})\sqrt{m_1+m_2/2}-\epsilon_{5}^{(g)}\sqrt{n})\leq 0\nonumber \\
& & f_i(\x)\leq 0,i=1,2,\dots,l. \label{eq:addprobanal7}
\end{eqnarray}
On the other hand one would have for a corresponding replacement of  (\ref{eq:ubprobanal5})
\begin{eqnarray}
U_{afc}^{(1)}=\min_{\x} & & f(\x)-\xi_D^{(u)}(f)\nonumber \\
\mbox{subject to} & & -\g^T\x+\|\x\|_2((1+\epsilon_{1}^{(m)})\sqrt{m_1+m_2/2}+\epsilon_{5}^{(g)}\sqrt{n})\leq 0\nonumber \\
& & f_i(\x)\leq 0,i=1,2,\dots,l. \label{eq:addprobanal8}
\end{eqnarray}
Of course, for all practical purposes programs (\ref{eq:addprobanal7}) and (\ref{eq:addprobanal8}) are basically equivalent. Statement of Lemma \ref{lemma:lblincons} would then remain in place with the only difference being that $L^{(1)}$ should be replaced by $L_{afc}^{(1)}$. Similarly, Lemma \ref{lemma:ublincons} would remain correct with $U^{(1)}$ being replaced by $U_{afc}^{(1)}$ and with an $f(\x)$ being such that the following modified version of (\ref{eq:ubrandlincons3}) holds
\begin{eqnarray}
\xi_{afc}(f,A,B) & = & \min_{\x}\max_{\lambda\geq 0,\gamma_i\geq 0,\nu} f(\x)+\nu^TA\x+\lambda^TB\x+\sum_{i=1}^{l}\gamma_i f_i(\x)\nonumber \\
& = & \max_{\lambda\geq 0,\gamma_i\geq 0,\nu}\min_{\x} f(\x)+\nu^TA\x+\lambda^TB\x+\sum_{i=1}^{l}\gamma_i f_i(\x). \label{eq:addubrandlincons3}
\end{eqnarray}
What we presented above is a generic scenario where all functions $f_i(\x)$ are assumed to be deterministic. Of course some of the additional constraints (sometimes even all of them) can be random functions as well. Then they typically can be massaged further, either when handling (\ref{eq:addprobanal7}) (or (\ref{eq:addprobanal8})) or in the derivation process from Section \ref{sec:randlincons}. However, the way to handle them is typically problem specific and we typically treat them on the individual case basis and choose to present such discussions elsewhere.

It is of course relatively easy to see that the non-homogenous case from the previous subsection and the case of additional functional constraints considered in this subsection can easily be merged. We of course skip rewriting this easy exercise. Instead in the following section we provide a specific example to demonstrate how the entire mechanism can be applied. Moreover, the example will be selected so that the mechanism works in its full capacity, i.e. with all assumptions being satisfied and both Lemma \ref{lemma:lblincons} and Lemma \ref{lemma:ublincons} being useful and essentially providing matching lower and upper bounds on the optimal value of the objective function.

\section{An example: homogeneous f(\x) with spherical bounding constraint}
\label{sec:examplehomf}

In this section we demonstrate how the mechanism from previous sections can be applied on a particular optimization problem. We start by assuming a specific type of the objective function. We will assume that $f(\x)$ is a homogeneous function. Namely, let $f_h(\x)$ be such that
\begin{equation}
f_h(a\x)=a^df_h(\x),\label{eq:exmhomf1}
\end{equation}
for any $a>0$ and a fixed $d>0$. Then we say that function $f_h(\x)$ is positive homogeneous of degree $d$. Then for all practical purposes the optimization problem (\ref{eq:lincons}) is useless. Basically, if there is a feasible $\x$ such that $f_h(\x)<0$ one can then keep multiplying such an $\x$ by a sequence of arbitrarily large increasing constants $a$ and no matter how small $d$ is the value of $f_h(\x)$ will eventually keep converging to $-\infty$. To help making problem (\ref{eq:lincons}) bounded we will add an origin encapsulating closed set to act as an additional bounding constraint. There is really no restriction as what this constraint needs to be. However, to facilitate concrete computations we will assume the most typical spherical constraint. One then has a reformulated version of (\ref{eq:lincons})
\begin{eqnarray}
\xi_{h}(f_h,A,B)=\xi_{afc}(f_h,f_1,A,B)=\min_{\x} & & f_h(\x)\nonumber \\
\mbox{subject to} & & A\x=0\nonumber \\
& & B\x\leq 0\nonumber \\
& & f_1(\x)=\|\x\|_2-1\leq 0. \label{eq:exmlincons}
\end{eqnarray}
Now, the mechanism of Section \ref{sec:lbrandlincons} can be used. A way to provide a lower bound based on such a mechanism is to determine a quantity $\xi_D^{(l)}(f)$ such that $L_{h}^{(1)}$ below is non-negative with overwhelming probability.
\begin{eqnarray}
L_{h}^{(1)}=\min_{\x} & & f(\x)-\xi_D^{(l)}(f)\nonumber \\
\mbox{subject to} & & \g^T\x+\|\x\|_2((1-\epsilon_{1}^{(m)})\sqrt{m_1+m_2/2}-\epsilon_{5}^{(g)}\sqrt{n})\leq 0\nonumber \\
& & \|\x\|_2\leq 1. \label{eq:exmprobanal7}
\end{eqnarray}
Of course, the larger $\xi_D^{(l)}(f)$ is the harder for $L_{h}^{(1)}$ to stay non-negative. So, roughly speaking, the best $\xi_D^{(l)}(f)$ would be the one that makes $L_{h}^{(1)}$ equal to zero (or to be more precise, the one that makes $L_{h}^{(1)}$ stay just above zero). When $\xi_h(f_h,A,B)<0$ The optimization problem in (\ref{eq:exmprobanal7}) can be simplified a bit
\begin{eqnarray}
L_{h}^{(1)}=\min_{\x} & & f(\x)-\xi_D^{(l)}(f)\nonumber \\
\mbox{subject to} & & \g^T\x+((1-\epsilon_{1}^{(m)})\sqrt{m_1+m_2/2}-\epsilon_{5}^{(g)}\sqrt{n})\leq 0\nonumber \\
& & \|\x\|_2\leq 1. \label{eq:exmprobanal8}
\end{eqnarray}
(Throughout the presentation in the rest of this section we pretty much ignore scenario when there is no $\x$ such that $\xi_h(f_h,A,B)<0$, since in that case one trivially has $\xi_h(f_h,A,B)=0$.) On the other hand, if one sets $f_1(\x)=\|\x\|_2-1$ and $f_h(\x)$ is such that (\ref{eq:addubrandlincons3}) holds then one can also utilize the mechanism of Section \ref{sec:ubrandlincons}. A way to provide an upper bound on $\xi_h(f_h,A,B)$ based on such a mechanism is to determine a quantity $\xi_D^{(u)}(f)$ such that $U_{h}^{(1)}$ below is non-positive with overwhelming probability.
\begin{eqnarray}
U_{h}^{(1)}=\min_{\x} & & f_h(\x)-\xi_D^{(u)}(f)\nonumber \\
\mbox{subject to} & & \g^T\x+\|\x\|_2((1+\epsilon_{1}^{(m)})\sqrt{m_1+m_2/2}+\epsilon_{5}^{(g)}\sqrt{n})\leq 0\nonumber \\
& & \|\x\|_2\leq 1. \label{eq:exmprobanal9}
\end{eqnarray}
Of course, the smaller $\xi_D^{(u)}(f)$ is the harder for $U_{h}^{(1)}$ to stay non-positive. Again, roughly speaking, the best $\xi_D^{(u)}(f)$ would be the one that makes $U_{h}^{(1)}$ equal to zero (or to be more precise, the one that makes $U_{h}^{(1)}$ stay just below zero). The optimization problem in (\ref{eq:exmprobanal9}) can be simplified a bit
\begin{eqnarray}
U_{h}^{(1)}=\min_{\x} & & f_h(\x)-\xi_D^{(u)}(f)\nonumber \\
\mbox{subject to} & & \g^T\x+((1+\epsilon_{1}^{(m)})\sqrt{m_1+m_2/2}+\epsilon_{5}^{(g)}\sqrt{n})\leq 0\nonumber \\
& & \|\x\|_2\leq 1. \label{eq:exmprobanal10}
\end{eqnarray}
Of course, roughly speaking (basically ignoring all $\epsilon$'s), for all practical purposes programs (\ref{eq:exmprobanal8}) and (\ref{eq:exmprobanal10}) are equivalent, which essentially means that if $f_h(\x)$ is such that (\ref{eq:addubrandlincons3}) holds then not only will $\xi_D^{(l)}(f)$ be a lower bound on $\xi_h(f_h,A,B)$ with probability $1$ as $n\rightarrow \infty$, but also its a small variation $\xi_D^{(u)}(f)$ will be an upper bound on $\xi_h(f_h,A,B)$ with probability $1$ as $n\rightarrow \infty$. Or in other words, the probability that $\xi_h(f_h,A,B)$ will substantially deviate away from $\xi_D^{(l)}(f)$ will go to zero as $n\rightarrow\infty$.

Now, to demonstrate how one would proceed further we will look at a couple of particular examples of homogeneous functions.

\subsection{Purely linear f(\x)}
\label{sec:exampurelin}

We will first look at quite likely the simplest possible example for $f(\x)$, namely a purely linear function. So, we will set
\begin{equation}
f_{lp}(\x)=\sum_{i=1}^{n}\x_i.\label{eq:exmpurlin1}
\end{equation}
Then (\ref{eq:exmprobanal8}) becomes
\begin{eqnarray}
L_{lp}^{(1)}=\min_{\x} & & \sum_{i=1}^{n}\x_i-\xi_D^{(l)}(f_{lp})\nonumber \\
\mbox{subject to} & & \g^T\x+((1-\epsilon_{1}^{(m)})\sqrt{m_1+m_2/2}-\epsilon_{5}^{(g)}\sqrt{n})\leq 0\nonumber \\
& & \|\x\|_2\leq 1. \label{eq:exmpurlinprobanal8}
\end{eqnarray}
Also, to make writing easier we will set
\begin{equation}
\sqrt{D^{(l)}}=((1-\epsilon_{1}^{(m)})\sqrt{\alpha_1+\alpha_2/2}-\epsilon_{5}^{(g)}).\label{eq:exmpurlin2}
\end{equation}
Now we rewrite (\ref{eq:exmpurlinprobanal8}) in the following more convenient way
\begin{eqnarray}
L_{lp}^{(1)}=\min_{\x}\max_{\lambda\geq 0} & & \sum_{i=1}^{n}\x_i+\lambda \g^T\x+\lambda\sqrt{D^{(l)}}\sqrt{n}-\xi_D^{(l)}(f_{lp})\nonumber \\
\mbox{subject to} & & \|\x\|_2\leq 1. \label{eq:exmpurlinprobanal9}
\end{eqnarray}
Since the duality easily holds one then further has
\begin{eqnarray}
L_{lp}^{(1)}=\max_{\lambda\geq 0}\min_{\x} & & \sum_{i=1}^{n}\x_i+\lambda \g^T\x+\lambda\sqrt{D^{(l)}}\sqrt{n}-\xi_D^{(l)}(f_{lp})\nonumber \\
\mbox{subject to} & & \|\x\|_2\leq 1. \label{eq:exmpurlinprobanal10}
\end{eqnarray}
After solving the inner minimization we finally have
\begin{equation}
L_{lp}^{(1)}=\max_{\lambda\geq 0} (-\|\1+\lambda \g^T\|_2+\lambda\sqrt{D^{(l)}}\sqrt{n})-\xi_D^{(l)}(f_{lp}), \label{eq:exmpurlinprobanal11}
\end{equation}
where $\1$ is the $n$-dimensional column vector of all ones. Now, clearly, $L_{lp}^{(1)}$ is a random quantity. To completely understand its random behavior one would need to study it in full detail. However, since this paper is mostly concerned with a conceptual approach rather than with the details of particular calculations we will skip all unnecessary portions and focus only on the main results. To that end we will just mention without proving that $L_{lp}^{(1)}$ concentrates around its mean with overwhelming probability (the proof of this fact is not hard; however we do feel that going into such details would sidetrack our exposition; instead we do mention that a great deal of details needed for proofs of this type can be found in e.g. \cite{StojnicCSetam09,StojnicUpper10} as well as in many general probability type of references). Given all of this it is clear that to apply results of Lemma \ref{lemma:lblincons} it is then enough to compute $EL_{lp}^{(1)}$ and then choose $\xi_D^{(l)}(f_{lp})$ such that $EL_{lp}^{(1)}\geq 0$. When $n$ is large one then has
\begin{equation}
\lim_{n\rightarrow\infty}\frac{EL_{lp}^{(1)}}{\sqrt{n}}=\max_{\lambda\geq 0} (-\sqrt{1+\lambda^2}+\lambda\sqrt{D^{(l)}})-\lim_{n\rightarrow\infty}\frac{\xi_D^{(l)}(f_{lp})}{\sqrt{n}}, \label{eq:exmpurlinprobanal12}
\end{equation}
which after solving over $\lambda$ gives
\begin{equation}
\lim_{n\rightarrow\infty}\frac{EL_{lp}^{(1)}}{\sqrt{n}}=
\begin{cases}-\sqrt{1-D^{(l)}}-\lim_{n\rightarrow\infty}\frac{\xi_D^{(l)}(f_{lp})}{\sqrt{n}}, & \mbox{if} \quad D^{(l)}\leq 1\\
-\lim_{n\rightarrow\infty}\frac{\xi_D^{(l)}(f_{lp})}{\sqrt{n}}, & \mbox{otherwise}
\end{cases}. \label{eq:exmpurlinprobanal13}
\end{equation}
Now if we recall on the definition of $D^{(l)}$ from (\ref{eq:exmpurlin2}) and set
\begin{equation}
\xi_D^{(l)}(f_{lp})=
\begin{cases}-\sqrt{1-((1-\epsilon_{1}^{(m)})\sqrt{\alpha_1+\alpha_2/2}-\epsilon_{5}^{(g)})^2}\sqrt{n}, & \mbox{if} \quad ((1-\epsilon_{1}^{(m)})\sqrt{\alpha_1+\alpha_2/2}-\epsilon_{5}^{(g)})^2\leq 1\\
0, & \mbox{otherwise}
\end{cases}, \label{eq:exmpurlinprobanal14}
\end{equation}
we then based on Lemma \ref{lemma:lblincons} and previous discussion have
\begin{equation}
\lim_{n\rightarrow\infty}P(\xi_h(f_{lp},A,B)>\xi_D^{(l)}(f_{lp}))=1,\label{eq:exmplbbxi}
\end{equation}
where $\xi_D^{(l)}(f_{lp})$ is as in (\ref{eq:exmpurlinprobanal14}).

Since for $f_{lp}(\x)$ from (\ref{eq:exmpurlin1}) and $f_1(\x)=\|\x\|_2-1$ (\ref{eq:addubrandlincons3}) holds, one can now, analogously to (\ref{eq:exmpurlin2}), set
\begin{equation}
\sqrt{D^{(u)}}=((1+\epsilon_{1}^{(m)})\sqrt{\alpha_1+\alpha_2/2}+\epsilon_{5}^{(g)}),\label{eq:exmpurlinub2}
\end{equation}
and write the following analogue to (\ref{eq:exmpurlinprobanal9})
\begin{eqnarray}
U_{lp}^{(1)}=\min_{\x}\max_{\lambda\geq 0} & & \sum_{i=1}^{n}\x_i-\lambda \g^T\x+\lambda\sqrt{D^{(u)}}\sqrt{n}-\xi_D^{(u)}(f_{lp})\nonumber \\
\mbox{subject to} & & \|\x\|_2\leq 1. \label{eq:exmpurlinprobanal9ub}
\end{eqnarray}
After repeating previous arguments and relying on Lemma \ref{lemma:ublincons} one then arrives at
\begin{equation}
\lim_{n\rightarrow\infty}P(\xi_h(f_{lp},A,B)<\xi_D^{(u)}(f_{lp}))=1,\label{eq:exmplbbxiub}
\end{equation}
where $\xi_D^{(u)}(f_{lp})$ would analogously to (\ref{eq:exmpurlinprobanal14}) be
\begin{equation}
\xi_D^{(u)}(f_{lp})=
\begin{cases}-\sqrt{1-((1+\epsilon_{1}^{(m)})\sqrt{\alpha_1+\alpha_2/2}+\epsilon_{5}^{(g)})^2}\sqrt{n}, & \mbox{if} \quad ((1+\epsilon_{1}^{(m)})\sqrt{\alpha_1+\alpha_2/2}+\epsilon_{5}^{(g)})^2\leq 1\\
0, & \mbox{otherwise}
\end{cases}. \label{eq:exmpurlinprobanal14ub}
\end{equation}

We summarize the above presentation in the following convenient lemma.

\begin{lemma}
Consider optimization problem in (\ref{eq:exmlincons}). Let $f_h(\x)=f_{lp}(\x)=\sum_{i=1}^{n}\x_i$. Let $A$ be an $m_1\times n$ matrix with i.i.d. standard normal components. Let $B$ be an $m_2\times n$ matrix with i.i.d. standard normal components. Assume that $n$ is large and that $m_1=\alpha_1 n$ and $m_2=\alpha_2 n$ where $\alpha_1$ and $\alpha_2$ are constants independent of $n$. Let $\xi_D^{(l)}(f_{lp})$ and $\xi_D^{u)}(f_{lp})$ be as in
(\ref{eq:exmpurlinprobanal14}) and (\ref{eq:exmpurlinprobanal14ub}), respectively. Let $\epsilon$'s in (\ref{eq:exmpurlinprobanal14}) and (\ref{eq:exmpurlinprobanal14ub}) be arbitrarily small constants independent of $n$. Then,
\begin{equation*}
\lim_{n\rightarrow\infty}P(\xi_D^{(l)}(f_{lp})<\xi_h(f_{lp},A,B)<\xi_D^{(u)}(f_{lp}))=1,\label{eq:exmpllemma}
\end{equation*}
\label{lemma:exmpl}
\end{lemma}
\begin{proof}
Follows from previous discussion.
\end{proof}

More informally, assume the setup of Lemma \ref{lemma:exmpl}. If $1-\alpha_1-\alpha_2/2\> 0$, one then has that with very low probability the optimal value of the objective function in (\ref{eq:exmlincons}), $\xi_h(f_{lp},A,B)$, would deviate from $-\sqrt{1-\alpha_1-\alpha_2/2}\sqrt{n}$. On the other hand if $1-\alpha_1-\alpha_2/2< 0$ then with very high probability the optimal value of the objective function in (\ref{eq:exmlincons}), $\xi_h(f_{lp},A,B)$, is zero.

\subsubsection{Numerical example}
\label{sec:numexmpurlin}

To give a bit more flavor as to how useful practically would be the results from the previous subsection, we conducted a limited set of numerical experiments. Namely, we solved problem (\ref{eq:exmlincons}) with $A$ and $B$ as randomly generated i.i.d. Gaussian matrices and $f_h(\x)=f_{lp}(\x)=\sum_{i=1}^{n}\x_i$. We repeated our experiment a number of times with different (but of course random) $A$ and $B$. The results we obtained are summarized in Table \ref{tab:exmpurlin}. The second row contains the numerical values obtained through the simulations and the third row contains the numerical values that the above theory predicts. As can be seen from Table \ref{tab:exmpurlin}, even for a fairly small value of $n$ one has a solid agreement between what the above theory predicts and the results obtained through numerical experiments. The results we presented in Table \ref{tab:exmpurlin} are given for the expected values whereas Lemma \ref{lemma:exmpl} gives a probabilistic type of behavior. However, as we mentioned earlier, all important quantities do concentrate and they do concentrate around their mean values.

\begin{table}
\caption{Experimental results for (\ref{eq:exmlincons}); $\alpha_1=0.5$; (\ref{eq:exmlincons}) was run $1000$ times with $n=200$}\vspace{.1in}
\hspace{-0in}\centering
\begin{tabular}{||c|c|c|c|c|c|c||}\hline\hline
 $\alpha_2$ & $0.5$ & $0.6$ & $0.7$ & $0.8$ & $0.9$ & $1$ \\ \hline\hline
 $\frac{E(L_{lp}^{(1)}+\xi_D^{(l)})}{\sqrt{n}} $ -- (\mbox{sim.}) 
 &  $-0.4979$ & $-0.4433$ & $-0.3792$ & $-0.3040$ & $-0.2044$ & $-0.0723$ \\ \hline
 $\lim_{n\rightarrow \infty}\frac{E(L_{lp}^{(1)}+\xi_D^{(l)})}{\sqrt{n}} $ -- (\mbox{th.}) 
 &  $-0.5000$ & $-0.4472$ & $-0.3873$ & $-0.3162$ & $-0.2236$ & $-0.0000$ \\ \hline\hline
\end{tabular}
\label{tab:exmpurlin}
\end{table}

\subsection{General linear f(\x)}
\label{sec:examgenlin}

We will now extend a bit the results from the previous subsection. Namely, instead of looking at a purely linear function $f(\x)$ we will look at general linear functions. So, we will set
\begin{equation}
f_{gl}(\x)=\sum_{i=1}^{n}\c_i\x_i=\c^T\x,\label{eq:exmgenlin1}
\end{equation}
where $\c$ is a deterministic (fixed) $n\times 1$ vector from $R^n$. For concreteness we will also set $C_{gl}=\frac{\|\c\|_2}{\sqrt{n}}$ and assume $C_{gl}<\infty$ as $n\rightarrow \infty$. As in previous subsection one can then consider
\begin{eqnarray}
L_{gl}^{(1)}=\min_{\x} & & \sum_{i=1}^{n}\c_i\x_i-\xi_D^{(l)}(f_{gl})\nonumber \\
\mbox{subject to} & & \g^T\x+((1-\epsilon_{1}^{(m)})\sqrt{m_1+m_2/2}-\epsilon_{5}^{(g)}\sqrt{n})\leq 0\nonumber \\
& & \|\x\|_2\leq 1. \label{eq:exmgenlinprobanal8}
\end{eqnarray}
After repeating all the steps from the previous subsection one then arrives at
\begin{equation}
\lim_{n\rightarrow\infty}\frac{EL_{gl}^{(1)}}{\sqrt{n}}=\max_{\lambda\geq 0} (-\sqrt{C_{gl}^2+\lambda^2}+\lambda\sqrt{D^{(l)}})-\lim_{n\rightarrow\infty}\frac{\xi_D^{(l)}(f_{gl})}{\sqrt{n}}, \label{eq:exmpurlinprobanal12}
\end{equation}
which after solving over $\lambda$ gives
\begin{equation}
\lim_{n\rightarrow\infty}\frac{EL_{lp}^{(1)}}{\sqrt{n}}=
\begin{cases}-C_{gl}\sqrt{1-D}-\lim_{n\rightarrow\infty}\frac{\xi_D^{(l)}(f_{lp})}{\sqrt{n}}, & \mbox{if} \quad D\leq 1\\
-\lim_{n\rightarrow\infty}\frac{\xi_D^{(l)}(f_{lp})}{\sqrt{n}}, & \mbox{otherwise}
\end{cases}. \label{eq:exmpurlinprobanal13}
\end{equation}
One can then repeat all remaing arguments from the previous subsection to arrive at the following (more general) analogue of Lema \ref{lemma:exmpl}.
\begin{lemma}
Consider optimization problem in (\ref{eq:exmlincons}). Let $f_h(\x)=f_{gl}(\x)=\sum_{i=1}^{n}\c_i\x_i$, where $\c$ is a deterministic (fixed) $n\times 1$ vector from $R^n$. Set $C_{gl}=\frac{\|\c\|_2}{\sqrt{n}}$ and assume $C_{gl}<\infty$ as $n\rightarrow \infty$. Let $A$ be an $m_1\times n$ matrix with i.i.d. standard normal components. Let $B$ be an $m_2\times n$ matrix with i.i.d. standard normal components. Assume that $n$ is large and that $m_1=\alpha_1 n$ and $m_2=\alpha_2 n$ where $\alpha_1$ and $\alpha_2$ are constants independent of $n$. Let $\xi_D^{(l)}(f_{gl})=C_{gl}\xi_D^{(l)}(f_{lp})$ and $\xi_D^{u)}(f_{gl})=C_{gl}\xi_D^{u)}(f_{lp})$ where $\xi_D^{(l)}(f_{lp})$ and $\xi_D^{u)}(f_{lp})$ are as in
(\ref{eq:exmpurlinprobanal14}) and (\ref{eq:exmpurlinprobanal14ub}), respectively. Let $\epsilon$'s in (\ref{eq:exmpurlinprobanal14}) and (\ref{eq:exmpurlinprobanal14ub}) be arbitrarily small constants independent of $n$. Then,
\begin{equation*}
\lim_{n\rightarrow\infty}P(\xi_D^{(l)}(f_{gl})<\xi_h(f_{gl},A,B)<\xi_D^{(u)}(f_{gl}))=1,\label{eq:exmgllemma}
\end{equation*}
\label{lemma:exmgl}
\end{lemma}
\begin{proof}
Follows from previous discussion.
\end{proof}

\noindent \textbf{Remark:} Knowing results of Lemma \ref{lemma:exmpl} one can deduce Lemma \ref{lemma:exmgl} even faster. For example, one can observe that $f_{gl}(x)=\c^T\x=C_{gl}\1^TQ_{\c}\x$ where $Q_{\c}$ is an $n\times n$ matrix such that $Q_{\c}^TQ_{\c}=I$. Then (\ref{eq:exmlincons}) with $f_h(\x)=f_{gl}(\x)$ becomes
\begin{eqnarray}
\xi_{h}(f_h,A,B)=\xi_{afc}(f_h,f_1,A,B)=\min_{\x} & & C_{gl}\1^TQ_{\c}\x\nonumber \\
\mbox{subject to} & & AQ_{\c}^TQ_{\c}\x=0\nonumber \\
& & BQ_{\c}^TQ_{\c}\x\leq 0\nonumber \\
& & f_1(\x)=\|Q_{\c}\x\|_2-1\leq 0. \label{eq:exmlinconsrot}
\end{eqnarray}
After a change of variables $A_{rot}=AQ_{\c}^T$, $B_{rot}=BQ_{\c}^T$, and $\x_{rot}=Q_{\c}\x$ one further has
\begin{eqnarray}
\xi_{h}(f_h,A,B)=\xi_{afc}(f_h,f_1,A,B)=\min_{\x_{rot}} & & C_{gl}\1^T\x_{rot}\nonumber \\
\mbox{subject to} & & A_{rot}\x_{rot}=0\nonumber \\
& & B_{rot}\x_{rot}\leq 0\nonumber \\
& & f_1(\x)=\|\x_{rot}\|_2-1\leq 0. \label{eq:exmlinconsrot1}
\end{eqnarray}
Now, observing that due to rotational invariance of Gaussian distribution matrices $A_{rot}$ and $B_{rot}$ are again comprised of i.i.d. standard normals one effectively has the same optimization problem as in the previous subsection. The only difference is that the objective function is multiplied by $C_{gl}$ which is exactly what Lemma \ref{lemma:exmgl} states should be the case.

\subsection{A more general homogeneous f(\x)}
\label{sec:exammg}

In this subsection we will look at a more general homogeneous function $f_h(\x)$. Namely, we will set
\begin{equation}
f_h(\x)=f_{bp}(\x)=\sum_{i=1}^{n-k}|\x_i|+\sum_{i=n-k+1}^{n}\x_i,\label{eq:exmmgf}
\end{equation}
where $k=\beta n$ and $\beta\leq \alpha_1$ is a constant independent of $n$. This function is an interesting choice for at least three reasons. First, it appears as a very important object in studying sparse solutions of random under-determined linear systems of equations. Second, it is a function for which (\ref{eq:addubrandlincons3}) holds. And third, it has a nice structure that allows one to actually analytically compute $\xi_D^{(l)}$ (and since (\ref{eq:addubrandlincons3}) holds then $\xi_D^{(u)}$ as well). We will below closely follow the presentation of Section \ref{sec:exampurelin}. To that end we start with (\ref{eq:exmprobanal8}) which in the case of interest here simplifies to (we are again mostly concern with the scenario where $\xi_h(f_h,A,B)=\xi_h(f_{bp},A,B)<0$)
\begin{eqnarray}
L_{bp}^{(1)}=\min_{\x} & & \sum_{i=1}^{n-k+1}|\x_i|+\sum_{i=n-k+1}^{n}\x_i-\xi_D^{(l)}(f_{bp})\nonumber \\
\mbox{subject to} & & \g^T\x+((1-\epsilon_{1}^{(m)})\sqrt{m_1+m_2/2}-\epsilon_{5}^{(g)}\sqrt{n})\leq 0\nonumber \\
& & \|\x\|_2\leq 1. \label{eq:exmmgprobanal8}
\end{eqnarray}
Also, to make writing easier, as in Section \ref{sec:exampurelin}, we will use
$\sqrt{D^{(l)}}$ from (\ref{eq:exmpurlin2}). Now we rewrite (\ref{eq:exmmgprobanal8}) in the following more convenient way
\begin{eqnarray}
L_{bp}^{(1)}=\min_{\x}\max_{\lambda\geq 0} & & \sum_{i=1}^{n-k+1}|\x_i|+\sum_{i=n-k+1}^{n}\x_i+\lambda \g^T\x+\lambda\sqrt{D^{(l)}}\sqrt{n}-\xi_D^{(l)}(f_{bp})\nonumber \\
\mbox{subject to} & & \|\x\|_2\leq 1. \label{eq:exmmgprobanal9}
\end{eqnarray}
Since the duality easily holds one then further has
\begin{eqnarray}
L_{bp}^{(1)}=\max_{\lambda\geq 0}\min_{\x} & & \sum_{i=1}^{n-k+1}|\x_i|+\sum_{i=n-k+1}^{n}\x_i+\lambda \g^T\x+\lambda\sqrt{D^{(l)}}\sqrt{n}-\xi_D^{(l)}(f_{bp})\nonumber \\
\mbox{subject to} & & \|\x\|_2\leq 1. \label{eq:exmmgprobanal10}
\end{eqnarray}
After solving the inner minimization we finally have
\begin{eqnarray}
L_{bp}^{(1)} & = & \max_{\lambda\geq 0} (-\sqrt{\|(\1_{n-k}-\lambda |\g_{1:n-k}|)_-\|_2^2+\|(\1_{k}+\lambda \g_{n-k+1:n}\|_2^2}+\lambda\sqrt{D^{(l)}}\sqrt{n})-\xi_D^{(l)}(f_{bp})\nonumber \\
& = & \max_{\theta>0} (\theta^{-1}(-\sqrt{\|(\theta\1_{n-k}- |\g_{1:n-k}|)_-\|_2^2+\|(\theta\1_{k}+ \g_{n-k+1:n}\|_2^2}+
\sqrt{D^{(l)}}\sqrt{n}))-\xi_D^{(l)}(f_{bp}), \nonumber \\\label{eq:exmmgprobanal11}
\end{eqnarray}
where $\1_{n-k}$ and $\1_{k}$ are the $n-k$- and $k$-dimensional column vectors of all ones respectively. Also, $\g_{1:n-k}$ and $\g_{n-k+1:n-k}$ are vectors comprised of first $n-k$ and last $k$ components of $\g$, respectively. Vector $(\1_{n-k}-\lambda |\g_{1:n-k}|)_-$ is a vector comprised only of negative components of vector $(\1_{n-k}-\lambda |\g_{1:n-k}|)$ and analogously vector $(\theta\1_{n-k}- |\g_{1:n-k}|)_-$ is a vector comprised only of negative components of vector $(\theta \1_{n-k}- |\g_{1:n-k}|)$..

Now, clearly, $L_{bp}^{(1)}$ is a random quantity. To completely understand its random behavior one would need to study it in full detail. However, since this paper is mostly concerned with a conceptual approach rather than with the details of particular calculations we will, as in Section \ref{sec:exampurelin}, skip all unnecessary portions and focus only on the main results. To that end we will just mention without proving that $L_{bp}^{(1)}$ concentrates around its mean with overwhelming probability (the proof of this fact needs some work but it is conceptually easy; a majority of the details needed for the proof can be found in e.g. \cite{StojnicCSetam09,StojnicUpper10}). Given all of this it is clear that to apply results of Lemma \ref{lemma:lblincons} it is then enough to compute $EL_{bp}^{(1)}$ and then choose $\xi_D^{(l)}(f_{bp})$ such that $EL_{bp}^{(1)}\geq 0$. When $n$ is large one then has
\begin{equation}
\lim_{n\rightarrow\infty}\frac{EL_{bp}^{(1)}}{\sqrt{n}}=\max_{\theta> 0} (\theta^{-1}(-\sqrt{\frac{2(1-\beta)}{\sqrt{2\pi}}\int_{-\infty}^{-\theta}(\theta+z)e^{-z^2/2}dz+\beta(1+\theta^2)}+\sqrt{D^{(l)}}))
-\lim_{n\rightarrow\infty}\frac{\xi_D^{(l)}(f_{bp})}{\sqrt{n}}. \label{eq:exmmgprobanal12}
\end{equation}
After solving the integral one further has
\begin{equation}
\lim_{n\rightarrow\infty}\frac{EL_{bp}^{(1)}}{\sqrt{n}}=\max_{\theta> 0}
(\theta^{-1}(-\sqrt{2(1-\beta)(-\frac{\theta e^{-\theta^2/2}}{\sqrt{2\pi}}+\frac{(\theta^2+1)}{2}\mbox{erfc}(\frac{\theta}{\sqrt{2}}))+\beta(1+\theta^2)}+\sqrt{D^{(l)}}))
-\lim_{n\rightarrow\infty}\frac{\xi_D^{(l)}(f_{bp})}{\sqrt{n}}. \label{eq:exmmgprobanal13}
\end{equation}
Let
\begin{equation}
\phi^{(l)}(\theta)=
(\theta^{-1}(-\sqrt{2(1-\beta)(-\frac{\theta e^{-\theta^2/2}}{\sqrt{2\pi}}+\frac{(\theta^2+1)}{2}\mbox{erfc}(\frac{\theta}{\sqrt{2}}))+\beta(1+\theta^2)}+\sqrt{D^{(l)}})),\label{eq:defphitheta}
\end{equation}
and
\begin{equation}
\hat{\theta}^{(l)}=\max_{\theta> 0}\phi^{(l)}(\theta).\label{eq:defhattheta}
\end{equation}
Then one has
\begin{equation}
\lim_{n\rightarrow\infty}\frac{EL_{bp}^{(1)}}{\sqrt{n}}=
\begin{cases}\phi^{(l)}(\hat{\theta}^{(l)})-\lim_{n\rightarrow\infty}\frac{\xi_D^{(l)}(f_{bp})}{\sqrt{n}}, & \mbox{if} \quad \max_{\theta>0}\phi^{(l)}(\theta)<0\\
-\lim_{n\rightarrow\infty}\frac{\xi_D^{(l)}(f_{bp})}{\sqrt{n}}, & \mbox{otherwise}
\end{cases}. \label{eq:exmmgprobanal14}
\end{equation}
Now we set
\begin{equation}
\xi_D^{(l)}(f_{bp})=
\begin{cases}\phi^{(l)}(\hat{\theta}^{(l)})\sqrt{n}, & \mbox{if} \quad \max_{\theta>0}\phi^{(l)}(\theta)<0\\
0, & \mbox{otherwise}
\end{cases}. \label{eq:exmmgprobanal15}
\end{equation}
We then based on Lemma \ref{lemma:lblincons} and previous discussion have
\begin{equation}
\lim_{n\rightarrow\infty}P(\xi_h(f_{bp},A,B)>\xi_D^{(l)}(f_{bp}))=1,\label{eq:exmbpbbxi}
\end{equation}
where obviously $\xi_D^{(l)}(f_{bp})$ is as in (\ref{eq:exmmgprobanal15}).

Since for $f_{bp}(\x)$ from (\ref{eq:exmpurlin1}) and $f_1(\x)=\|\x\|_2-1$ (\ref{eq:addubrandlincons3}) holds, one can now make use of Lemma \ref{lemma:ublincons} to in a way upper-bound $\xi_h(f_{bp},A,B)$. One starts with writing the following analogue to (\ref{eq:exmpurlinprobanal9})
\begin{eqnarray}
U_{bp}^{(1)}=\min_{\x}\max_{\lambda\geq 0} & & \sum_{i=1}^{n-k}|\x_i|+\sum_{i=n-k+1}^{n}\x_i+\lambda \g^T\x+\lambda\sqrt{D^{(u)}}\sqrt{n}-\xi_D^{(u)}(f_{bp})\nonumber \\
\mbox{subject to} & & \|\x\|_2\leq 1. \label{eq:exmmgprobanal9ub}
\end{eqnarray}
After repeating previous arguments and relying on Lemma \ref{lemma:ublincons} one then arrives at
\begin{equation}
\lim_{n\rightarrow\infty}P(\xi_h(f_{bp},A,B)<\xi_D^{(u)}(f_{bp}))=1,\label{eq:exmmgbbxiub}
\end{equation}
where $\xi_D^{(u)}(f_{bp})$ would analogously to (\ref{eq:exmmgprobanal15}) be
\begin{equation}
\xi_D^{(u)}(f_{bp})=
\begin{cases}\phi^{(u)}(\hat{\theta})\sqrt{n}, & \mbox{if} \quad \max_{\theta>0}\phi^{(u)}(\theta)<0\\
0, & \mbox{otherwise}
\end{cases}, \label{eq:exmmgprobanal15ub}
\end{equation}
$D^{(u)}$ would be as in (\ref{eq:exmpurlinub2}), and $\phi^{(u)}(\theta)$  and $\hat{\theta}^{(u)}$ would analogously to (\ref{eq:defphitheta}) and (\ref{eq:defhattheta}) be
\begin{equation}
\phi^{(u)}(\theta)=
(\theta^{-1}(-\sqrt{2(1-\beta)(-\frac{\theta e^{-\theta^2/2}}{\sqrt{2\pi}}+\frac{(\theta^2+1)}{2}\mbox{erfc}(\frac{\theta}{\sqrt{2}}))+\beta(1+\theta^2)}+\sqrt{D^{(u)}})),\label{eq:defphithetaub}
\end{equation}
and
\begin{equation}
\hat{\theta}^{(u)}=\max_{\theta> 0}\phi^{(u)}(\theta).\label{eq:defhatthetaub}
\end{equation}

We summarize the above presentation in the following convenient lemma.

\begin{lemma}
Consider optimization problem in (\ref{eq:exmlincons}). Let $f_h(\x)=f_{bp}(\x)=\sum_{i=1}^{n-k}|\x_i|+\sum_{i=n-k+1}^{n}\x_i$. Let $A$ be an $m_1\times n$ matrix with i.i.d. standard normal components. Let $B$ be an $m_2\times n$ matrix with i.i.d. standard normal components. Assume that $n$ is large and that $m_1=\alpha_1 n$ and $m_2=\alpha_2 n$ where $\alpha_1$ and $\alpha_2$ are constants independent of $n$. Let $D^{(l)}$, $D^{(u)}$, $\phi^{(l)}(\theta)$, $\hat{\theta}^{(l)}$, $\phi^{(u)}(\theta)$, and $\hat{\theta}^{(u)}$ be as in (\ref{eq:exmpurlin2}), (\ref{eq:exmpurlinub2}), (\ref{eq:defphitheta}), (\ref{eq:defhattheta}), (\ref{eq:defphithetaub}), and (\ref{eq:defhatthetaub}), respectively. Further, let $\xi_D^{(l)}(f_{bp})$ and $\xi_D^{(u)}(f_{bp})$ be as in
(\ref{eq:exmmgprobanal15}) and (\ref{eq:exmmgprobanal15ub}), respectively. Let $\epsilon$'s in (\ref{eq:exmpurlin2}) and (\ref{eq:exmpurlinub2}) be arbitrarily small constants independent of $n$. Then,
\begin{equation*}
\lim_{n\rightarrow\infty}P(\xi_D^{(l)}(f_{bp})<\xi_h(f_{bp},A,B)<\xi_D^{(u)}(f_{bp}))=1,\label{eq:exmmglemma}
\end{equation*}
\label{lemma:exmmg}
\end{lemma}
\begin{proof}
Follows from previous discussion.
\end{proof}

\noindent \textbf{Remark:} Taking functional equation $\phi_{(l)}(\hat{\theta}^{(l)})$ (or $\phi_{(u)}(\hat{\theta}^{(u)})$) and equalling it with zero would give the critical dependence for $\beta$, $\alpha_1$, and $\alpha_2$ so that (\ref{eq:exmlincons}) has negative optimal value of the objective function with probability that goes to $1$ as $n\rightarrow\infty$. In fact, this (with $\alpha_2\rightarrow 0$) is precisely what was done in \cite{StojnicCSetam09,StojnicUpper10} to obtain the critical threshold
for success of $\ell_1$ optimization in recovering sparse solutions of random under-determined linear systems of equations (of course in \cite{StojnicCSetam09,StojnicUpper10} we were strictly interested in characterizing the critical threshold and properties of (\ref{eq:exmmgprobanal8}) in an as explicit way as possible and conducted a substantial further massage of (\ref{eq:defphithetaub}) and (\ref{eq:defhatthetaub}) which we clearly skip here).

\subsubsection{Numerical example}
\label{sec:numexmmg}

As in Subsection \ref{sec:exampurelin}, to give a bit more flavor as to how useful practically would be the results from the previous subsection, we conducted a limited set of numerical experiments. Namely, we solved problem (\ref{eq:exmlincons}) with $A$ and $B$ as randomly generated i.i.d. Gaussian matrices and $f_h(\x)=f_{bp}(\x)=\sum_{i=1}^{n-k}|\x_i|+\sum_{i=n-k+1}^{n}\x_i$. We again repeated our experiment a number of times with different (but of course random) $A$ and $B$. The results we obtained are summarized in Table \ref{tab:exmmg}. The second row contains the numerical values obtained through the simulations and the third row contains the numerical values that the above theory predicts. As can be seen from Table \ref{tab:exmmg}, even for a fairly small value of $n$, one as in Subsection \ref{sec:exampurelin}, has a solid agreement between what the above theory predicts and the results obtained through numerical experiments. The results we presented in Table \ref{tab:exmmg} are given for the expected values whereas Lemma \ref{lemma:exmmg} gives a probabilistic type of behavior. However, as we mentioned earlier, all important quantities do concentrate and they do concentrate around their mean values.

\begin{table}
\caption{Experimental results for (\ref{eq:exmlincons}); $\alpha_1=0.5$, $\alpha_2=0.5$; (\ref{eq:exmlincons}) was run $1000$ times with $n=200$}\vspace{.1in}
\hspace{-0in}\centering
\begin{tabular}{||c|c|c|c|c|c|c|c||}\hline\hline
 $\beta$ & $0.42$ & $0.5$ & $0.6$ & $0.7$ & $0.8$ & $0.9$ & $1$ \\ \hline\hline
 $\frac{E(L_{bp}^{(1)}+\xi_D^{(l)})}{\sqrt{n}} $ -- (\mbox{sim.}) 
 &  $-0.0265$ &  $-0.0904$ & $-0.1797$ & $-0.2645$ & $-0.3470$ & $-0.4242$ & $-0.4979$ \\ \hline
 $\lim_{n\rightarrow \infty}\frac{E(L_{bp}^{(1)}+\xi_D^{(l)})}{\sqrt{n}} $ -- (\mbox{th.}) 
 &  $-0.0189$ &  $-0.0936 $ & $-0.1825$ & $-0.2672$ & $-0.3481$ & $-0.4256 $ & $-0.5000$ \\ \hline\hline
\end{tabular}
\label{tab:exmmg}
\end{table}

\section{Conclusion}
\label{sec:conc}

In this paper we looked at classic linearly constrained optimization problems. We viewed them in a statistical context. We provided a general way of characterizing their optimal values. More specifically, we provided a generic strategy that can help create a lower-bound on the optimal value of the objective function. The strategy is based on transforming the original problem to its a simpler probabilistic alternate. On the other hand for a specific type of objective function we were then able to create an analogous strategy that can help create an upper-bound on the optimal value of the objective function. Moreover, probabilistically speaking the two bounds match which essentially means that the lower-bounding strategy (which works for any objective function) in certain scenarios is actually good enough to optimally characterize the entire problem.

We then mentioned that the presented framework is fairly powerful and presented ways how one can modify it to cover various other optimization problems. Still, the modifications that we presented are fairly simple and we chose to present them just to give an idea how relatively easy is to use the presented strategies. Of course a whole lot more can be done, i.e. the class of optimization problems where the strategies presented here will work is much wider then a few examples that we presented. However, since this is an introductory paper where we intended just to present the core concepts of a much bigger theory we skipped a detail discussion as to what the limits of our propositions are. Also, many of further modifications/extensions are typically problem specific and we thought that it is better to cover them separately and present such a coverage elsewhere.

What is also important to stress is that we viewed optimization problems in a statistical context. To be more precise, we assumed a typical Gaussian scenario where all random quantities in any of our problems are assumed to be i.i.d. standard normals. These assumptions substantially simplified the exposition but are not really necessary. In fact, all results presented here would actually hold for a fairly large class of random distributions. Proving that is not that hard. In fact there are many ways how it can be done, but typically would boil down to repetitive use of the central limit theorem. For example, a particularly simple and elegant approach would be the one of Lindeberg \cite{Lindeberg22}. Adapting our exposition to fit into the framework of the Lindeberg principle is relatively easy and in fact if one uses the elegant approach of \cite{Chatterjee06} pretty much a routine. Since we did not create these techniques we chose not to do these routine generalizations. However, to make sure that the interested reader has a full grasp of generality of the results presented here, we do emphasize again that pretty much any distribution that can be pushed through the Lindeberg principle would work in place of the Gaussian one that we used.

Since the theory that we presented above in a way establishes a random duality we decided to call it that way. Along the lines of the above mentioned probabilistic generality of our theory, we then coined the term \emph{regularly random duality} where under \emph{regularly random} we essentially view any randomness that eventually in large dimensional settings boils down to Gaussian. It is quite possible that there are other classes of randomness for which similar theories can be built. While they may not be as powerful as the Gaussian one it would certainly (at least from a mathematical point of view) be interesting to see what their shapes and forms are.

\begin{singlespace}
\bibliographystyle{plain}
\bibliography{RegRndDlt}
\end{singlespace}

\end{document}